\documentclass[fleqn,10pt]{wlscirep}
\usepackage[utf8]{inputenc}
\usepackage[T1]{fontenc}
\usepackage[utf8]{inputenc}
\usepackage{newunicodechar}
\newunicodechar{́}{'}  
\usepackage{textcomp}
\usepackage{lineno}
\usepackage{amsmath}
\usepackage{bm}
\usepackage{booktabs}
\usepackage{multirow}

\title{Optical hopfions with arbitrary two winding numbers}

\author[1,*]{Xinji Zeng}
\author[1,*]{Jinwen Wang}
\author[2,*]{Yun Chen}
\author[1]{Guang Liu}
\author[3]{Zhenyu Guo}
\author[1]{Yongkun Zhou}
\author[1]{Xin Yang}
\author[1]{Chengyuan Wang}
\author[1]{Dong Wei}
\author[1]{Haixia Chen}
\author[3,4,$\dagger$]{Yijie Shen}
\author[5,$\ddagger$]{Andrew Forbes}
\author[1,6,$\S$]{Hong Gao}
\affil[1]{Ministry of Education Key Laboratory for Nonequilibrium Synthesis and Modulation of Condensed Matter, Shaanxi Province Key Laboratory of Quantum Information and Quantum Optoelectronic Devices, School of Physics, Xi'an Jiaotong University, Xi'an 710049, China}
\affil[2]{Department of Physics, Huzhou University, Huzhou 313000, China}
\affil[3]{Center for Disruptive Photonic Technologies, School of Physical and Mathematical Sciences $\And$ The Photonics Institute, Nanyang Technological University, Singapore 637371, Singapore}
\affil[4]{School of Electrical and Electronic Engineering, Nanyang Technological University, Singapore, 639798, Singapore}
\affil[5]{School of Physics, University of the Witwatersrand, Johannesburg, South Africa.}
\affil[6]{State Key Laboratory of Human-Machine Hybrid Augmented Intelligence, Xi’an Jiaotong University, Xi’an 710049,
China}

\affil[$\dagger$]{Yijie.shen@ntu.edu.sg}
\affil[$\ddagger$]{andrew.forbes@wits.ac.za}
\affil[$\S$]{honggao@xjtu.edu.cn}
\affil[*]{These authors contributed equally to this work.}

\begin{abstract}
Hopfions, as three-dimensional topologically nontrivial structures described by poloidal and toroidal winding numbers, hold promise as robust information carriers in spintronics, functional materials, and optical communications. Although they have been experimentally realized in various physical systems, such realizations have been restricted to low orders, with the winding numbers lacking tunability. Here, using optical fields as our platform, we outline how to make tunable hopfions in any order with any winding number.  We use tailored superpositions of Laguerre-Gaussian modes in free-space as our construction, achieving effective control for arbitrary-order poloidal and toroidal winding numbers, which we demonstrate up to orders 5 and 3, respectively, for a new state-of-the-art.  The resulting torus-knot structures are visualized experimentally via polarization filaments, confirming the designed topological textures. Our work reports an exotic optical topologies observed in free space, provides a systematic route hopfions of any order, with implications for topological photonics, optical communications, and analogies in magnetic and condensed-matter systems.

\end{abstract}
\begin{document}

\flushbottom
\maketitle
%
%
\thispagestyle{empty}


\section*{Introduction}
Topological solitons, as quasiparticle structures whose topologically protected properties remain invariant under continuous deformation, have attracted extensive research interest across many branches of physics \cite{beach2005dynamics,parkin2008magnetic,mcgilly2015controlling,meier2016observation,veenstra2024non}. Among them, skyrmions in two-dimensional (2D) systems \cite{piette1995dynamics} are the most prominent, describing a mapping from the 2D sphere $S^2$ to the 2D plane $\mathbb{R}^2$, with their topological stability guaranteed by the homotopy group $\pi_2(S^2) = \mathbb{Z}$. Initially proposed in the Skyrme model to describe pion field configurations of protons and neutrons \cite{skyrme1962unified}, they are now extensively studied in fields like condensed matter physics and optics due to their potential as ideal carriers for next-generation high-capacity communication devices \cite{muhlbauer2009skyrmion,yu2010real,fert2017magnetic,tsesses2018optical,shen2024optical,zeng2025tailoring,wang2025generation}. For the three-dimensional (3D) case, mathematically, there exists a mapping from the 3D sphere $S^3$ to the 2D sphere $S^2$, known as the Hopf fibration. Each point on $S^2$ has a preimage that is a circle $S^1$, and these circles are interwoven to form a fiber bundle. Similarly, the homotopy group $\pi_3(S^2) = \mathbb{Z}$ ensures its topological stability. Based on this mapping, Faddeev et al. demonstrated the existence of a class of stable topological soliton solutions in a Lagrangian similar to the Skyrme model \cite{faddeev1976some}. These solutions, named hopfions, have their topological charge described by the Hopf invariant \cite{whitehead1947expression}. As quasi-particles in real 3D space, hopfions with nonzero topological charge necessarily form interlinked fibration structures through their field lines, representing a direct application of mathematics in physics. As a result, hopfions have been realized in high-energy physics \cite{faddeev1997stable}, magnets and ferroelectrics \cite{liu2018binding,rybakov2022magnetic,guslienko2024magnetic,zheng2023hopfion,sallermann2023stability,kent2021creation,luk2020hopfions}, Bose-Einstein condensates \cite{zou2021formation}, liquid crystals \cite{ackerman2015self,leask2025topological}, and cosmology \cite{cruz2007cosmic}. In particular, magnetic hopfions, with their small size and ability to be driven by electric currents and magnetic fields \cite{wang2019current,raftrey2021field,liu2020three}, also exhibit the topological Hall effect \cite{saji2023hopfion}, holding promise as topologically stable 3D information carriers.

For hopfions, the linking number between fibers is typically described by a pair of winding numbers: the poloidal winding number $p$ and the toroidal winding number $q$ \cite{kobayashi2014torus}. In the most extensively studied magnetic hopfions, only basic-order have been realized so far, such as $(p,q)=(1,1)$, due to the constraint of energy minimization on the total free energy of the system. Conversely, the optical field is limited by the Maxwell equations, offering greater flexibility in generating hopfions. The first experimental realization of optical hopfions in free space was achieved using the Stokes vector \cite{sugic2021particle}, and subsequent studies have produced optical hopfions with varied topological types and orders beyond the basic one \cite{shen2023topological,ehrmanntraut2023optical,droop2023transverse,lin2025space}. In addition to the Stokes vector, alternative approaches based on the spin \cite{wang2023photonic,wu2025photonic} or the phase \cite{wan2022scalar,lyu2024formation} of the photon have also been realized. These optical hopfions all utilize the longitudinal orbital angular momentum (OAM) of beams to realize the generation of $p$-th order. However, the highest poloidal winding number reported to date remain limited to $p=3$ \cite{shen2023topological,zhong2024toroidal}, and lacks tunability. Since transverse OAM has not been proposed in the spatial dimension under paraxial conditions thus far, the independent control of the toroidal winding number $q$ has not yet been demonstrated. While high-order vortex knots \cite{dennis2010isolated,tsvetkov2025sculpting,pires2025stability} and polarization knots \cite{larocque2018reconstructing} can be created in free space from singularities of structured light, these structures do not belong to the category of hopfions as the absence of a Hopf mapping. So far, a general method to generate hopfions with arbitrary tunable winding numbers in any physical system remains elusive.

Here, we propose and experimentally demonstrate a versatile approach to creating optical hopfions with tunable winding numbers in free space. By tailoring the complex amplitude profiles of two orthogonal polarization components within a vector beam, specifically through delicate manipulation of the Gouy phase instead of directly addressing transverse OAM, we construct polarization-defined filaments with programmable winding topology in 3D free space. Throughout propagation, the polarization distribution follows a Hopf map onto the fundamental Poincaré sphere. We experimentally achieve independent and combined control of the poloidal and toroidal winding numbers and get the corresponding Hopf invariants. Owing to its systematic design strategy, our method can be naturally extended to arbitrarily high orders. These results significantly expand the family of realizable optical topological solitons and provide a feasible route toward generating arbitrary-order hopfions in other physical systems, with potential implications for high-capacity optical information processing \cite{zhang2023magnon}, 3D spintronics \cite{fernandez2017three,gubbiotti20252025}, as well as light-matter interactions \cite{tamura2024three,wang2024measuring,zeng2024spatial,wang2020vectorial}.

\section*{Results}
\subsection*{Concepts of arbitrary-order hopfions} 
The topological textures of optical polarization states are usually described by the Stokes vector $\mathbf{S}=(S_x,S_y,S_z)$. As a polarization configuration in 3D space, the analytical expression for hopfions with $(p, q)$ winding numbers is (see supplementary material):
\begin{equation}\label{eq8}\begin{split}
    S_z&=P\frac{1-\cosh^{2p}{(\eta)}\tanh^{2q}{(\eta)}}{1+\cosh^{2p}{(\eta)}\tanh^{2q}{(\eta)}},\\
    S_x+iS_y&=\sqrt{1-S_z^2}\exp{[i(p\theta+q\phi)]},
    \end{split}
\end{equation}
where $P=\pm1$ is the polarity and we have introduced the toroidal coordinates $\mathbf{r}(\eta, \theta, \phi)$, which are related to the cylindrical coordinates $(\rho, \phi, z)$ as: $\rho=a\frac{\sinh{(\eta)}}{\cosh{(\eta)}-\cos{(\theta)}},\phi=\phi,z=a\frac{\sin{(\theta)}}{\cosh{(\eta)}-\cos{(\theta)}}$, $a$ is a scale parameter. For a given Stokes vector $\mathbf{S}$ on a hopfion, the isopolarization line (or filaments) form a torus knot in 3D space, also described by the winding numbers $(p, q)$. When $\eta = \eta_0$ is constant and $\theta = pt$, $\phi = qt$ with $t \in [0, 2\pi]$, the parametric equations for the torus knot can be obtained:
\begin{equation}\label{eq9}
    \begin{split}
        x(t)&=[R+r\frac{\cos{(pt)}}{R/r-\cos{(pt)}}]\cos{(qt)},
        \\y(t)&=[R+r\frac{\cos{(pt)}}{R/r-\cos{(pt)}}]\sin{(qt)},\\
        z(t)&=ar\frac{\sin{(pt)}}{R/r-\cos{(pt)}},
    \end{split}
\end{equation}
where we have used $x=\rho\cos{(\phi)},y=\rho\sin{(\phi)}$, the toroidal radius $R=\coth{(\eta_0)}$, the poloidal radius $r=1/\sinh{(\eta_0)}$. In addition to the winding numbers, the topological invariants of hopfions are generally defined by the Hopf index $N_H$:
\begin{equation}\label{eq10}
    N_H=\frac{1}{(4\pi)^2}\iiint\mathbf{B}\cdot\mathbf{A}\mathrm{d}x\mathrm{d}y\mathrm{d}z,
\end{equation}
where $B_i=\varepsilon_{i,j,k}\mathbf{S}\cdot(\partial_j\mathbf{S}\times\partial_k\mathbf{S})/2$, in which $\{i,j,k\}=\{x,y,z\}$, $\varepsilon$ is the Levi-Civita tensor, \(\mathbf{A}\) is the gauge vector potential such that $\nabla \times \mathbf{A} = \mathbf{B}$. For an ideal hopfion with winding number $(p, q)$, the Hopf index satisfies $N_H = pq$ \cite{kobayashi2014torus}.

\begin{figure*}[ht!]
    \vspace*{0mm} 
        \centering
        \includegraphics[height=7cm]{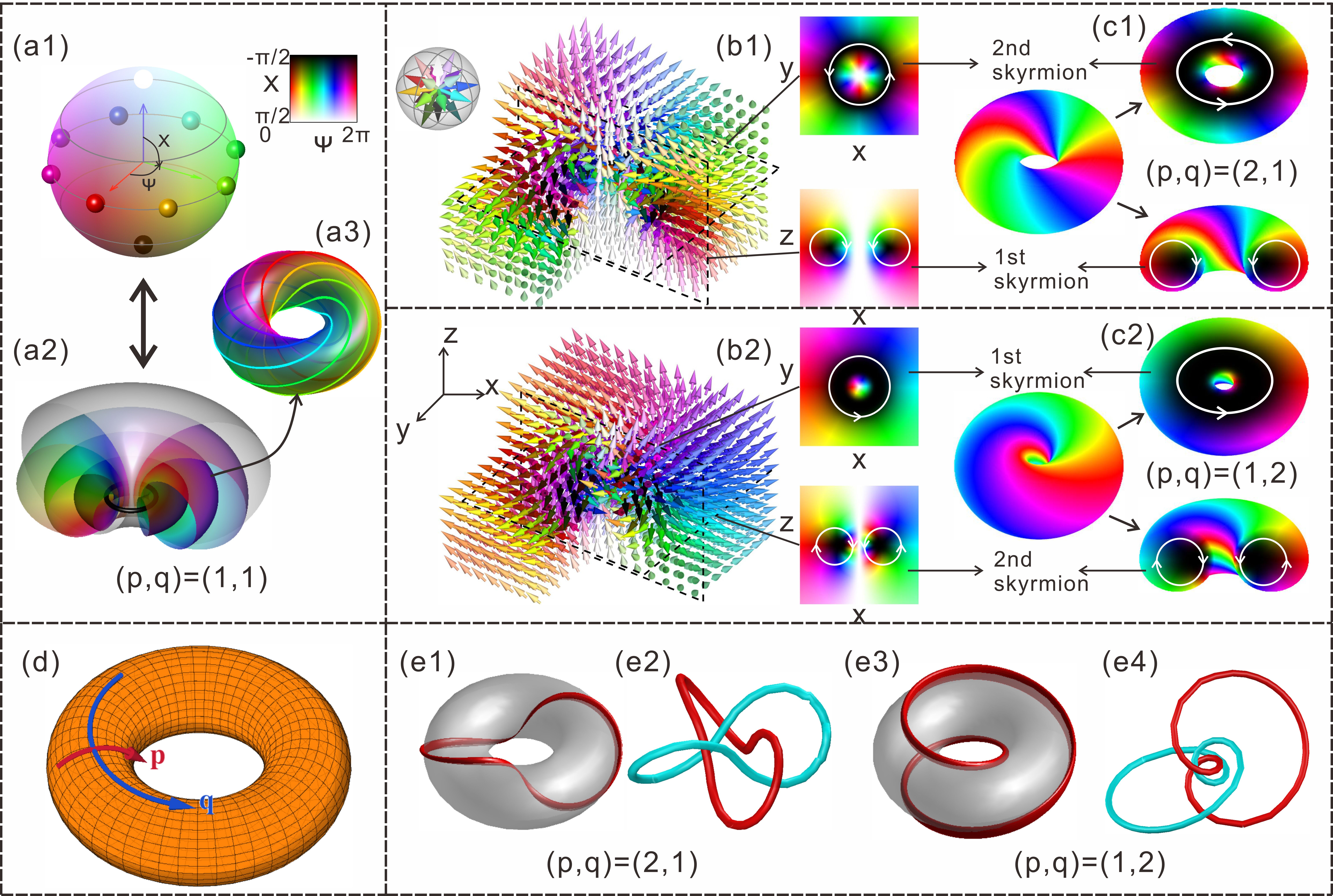}
        \\[0mm] 
        \caption{Schematic diagrams of the Hopf map, the structures of hopfions and torus knots. (a1) is the first-order Poincar\'e sphere with several polarization states marked by small spheres, the color map is based on the sphere's azimuthal angle $\psi$ and spatial angle $\chi$. (a2) shows the nested torus structure corresponding to the $S_z$ isosurface, with the torus at $S_z = 0$ displayed in (a3) and several torus knots drawn. (b1) and (b2) show the distribution of Stokes vectors for $(p, q) = (2, 1)$ and $(1, 2)$, respectively, with the polarization distributions in the $xy$ and $xz$ planes provided on the right. (c1) and (c2) show the solid torus structures of the polarizations with $S_z \leq 0$ corresponding to the winding numbers in (b1) and (b2), respectively, with their $xy$ and $xz$ cross-sections on the right. (d) indicates the poloidal and toroidal winding directions, labeled by $p$ and $q$. (e1) and (e3) are the torus knots for $(p,q)=(2,1)$ and $(1,2)$, with (e2) and (e4) show the interlinked orthogonal isopolarization lines of the two corresponding hopfions.
} 
        \label{fig1}
    \end{figure*}
    For a clearer conceptual illustration, we first provide an explanation of the basic-order hopfion. A schematic diagram of the Hopf map with winding number $(p, q) = (1, 1)$ is shown in Fig.\ref{fig1} (a1)-(a3). Fig.\ref{fig1} (a1) illustrates the first-order Poincaré sphere, which maps all polarization states to different hue colors and lightness according to the azimuthal angle $\psi$ and the spatial angle $\chi$. For example, when $(\psi, \chi) = (0, 0)$, the corresponding Stokes vector is $(S_x, S_y, S_z) = (1, 0, 0)$, representing horizontal linear polarization, indicated by red with a lightness of $1/2$. When $S_z$ is constant, it corresponds to a latitude line on the Poincaré sphere, which includes all hue colors. Through the Hopf map, this latitude line corresponds to a torus in 3D physical space, as shown in Fig.\ref{fig1} (a2). The south pole with $S_z = -1$ corresponds to the smallest torus, and as $S_z$ increases, larger nested tori are formed, with the north pole at $S_z = 1$ corresponding to the largest outer torus. On each torus, the isopolarization lines with constant $(S_x, S_y)$ form torus knots. The winding numbers $(p,q)$ describe how many times a knot winds around the major circle and the minor circle of the torus. Here the major circle refers to the circle around the center hole of the torus, while the minor circle refers to the circle around the cross-section of the torus tube. For example, the $(p, q) = (1, 1)$ torus knot shown in Fig.\ref{fig1} (a3) winds the torus once around the major circle and once around the minor circle, and thus it is a ring with $S_z = 0$. These torus knots of different colors are marked with small spheres of corresponding colors on the sphere in Fig.\ref{fig1} (a1). The entire process vividly illustrates the Hopf map $S^1 \hookrightarrow S^3 \xrightarrow{\pi} S^2$. 

Fig.\ref{fig1} (b1)-(b2) illustrate the 3D Stokes vector distributions of high-order hopfions. Fig.\ref{fig1} (b1) shows a second-order hopfion with winding number $(p, q) = (2, 1)$, and its polarization distributions in the $xy$ plane at $z = 0$ and the $xz$ plane at $y = 0$ are shown on the right. Here, $p = 2$ indicates that the polarization winds twice around the cross-section of the toroidal tube, as seen in the $xy$ plane with a marked skyrmion structure. The toroidal winding number $q = 1$ means the polarization winds once around the toroidal direction, marked in the $xz$ plane with a first-order skyrmion-anti-skyrmion pair structure. Fig.\ref{fig1} (b2) shows another second-order hopfion with winding number $(p, q) = (1, 2)$. It has a first-order skyrmion in the $xy$ plane and a second-order skyrmion-anti-skyrmion pair structure in the $xz$ plane with polarizations repeating twice, corresponding to $q = 2$. Thus, a qualitative criterion for identifying a hopfion with winding number $(p, q)$ is having a $p$-th order skyrmion in the $xy$ plane and a $q$-th order skyrmion-anti-skyrmion  pair in the $xz$ plane. Topology allows for variations in shape; for example, the $q$-th order skyrmion in the $xz$ plane can split into $q$ first-order skyrmions, as will be shown later.

Fig.\ref{fig1} (c1) and (c2) show the solid torus structures of the second-order hopfions with winding numbers $(p, q) = (2, 1)$ and $(1, 2)$ when $S_z \leq0$, respectively. The differences between the two second-order hopfions can be clearly seen from the $xy$ and $xz$ cross-sections of the tori. Fig.\ref{fig1} (e1) and (e3) show the torus knots with $(p, q) = (2, 1)$ and $(1, 2)$, respectively. When the parameter $p=2$, the torus knot threads through the central hole of the torus twice, whereas $q = 2$ indicates that the knot wraps around the core of the torus twice-once on the inner side and once on the outer. According to Eq.\eqref{eq8}, Fig.\ref{fig1} (e2) and (e4) show the isopolarization lines corresponding to the two hopfions with $\mathbf{S} = (1, 0, 0)$ and $(-1, 0, 0)$, respectively. They are consistent with the torus knot structure. The orthogonal isopolarization lines are linked with each other by the product of their winding numbers $pq$.

\subsection*{Construction of (p,1)-order optical hopfions}

To construct optical field modes that can propagate in free space, it is necessary to ensure that they are solutions to Maxwell's equations. As a particular solution to the scalar Helmholtz equation, the Laguerre–Gaussian (LG) mode $\mathrm{LG}_p^l$ is a commonly chosen candidate, with radial index $p$ and azimuthal index $l$, carrying an OAM of $l\hbar$ per photon \cite{zhan2009cylindrical}. For the paraxial vector Helmholtz equation, the solution is a vector beam expressed as $\mathbf{E} = E_R \mathbf{e_R} + E_L \mathbf{e_L}$, where $E_R$ and $E_L$ are the right- and left-circularly polarized (RCP and LCP) components of the beam. We can obtain hopfions in free space based on the Stokes vector if $E_R(E_L)$ is constructed by linearly superimposing LG beams of different orders. Experimentally, we utilize complex amplitude modulation based on a spatial light modulator (SLM) combined with digital propagation technology to design the required RCP and LCP components. The target vector beam is obtained through beam splitting and combining by a pair of beam displacer (see supplementary).

\begin{figure*}[ht!]
    \vspace*{0mm} 
        \centering
        \includegraphics[height=13cm]{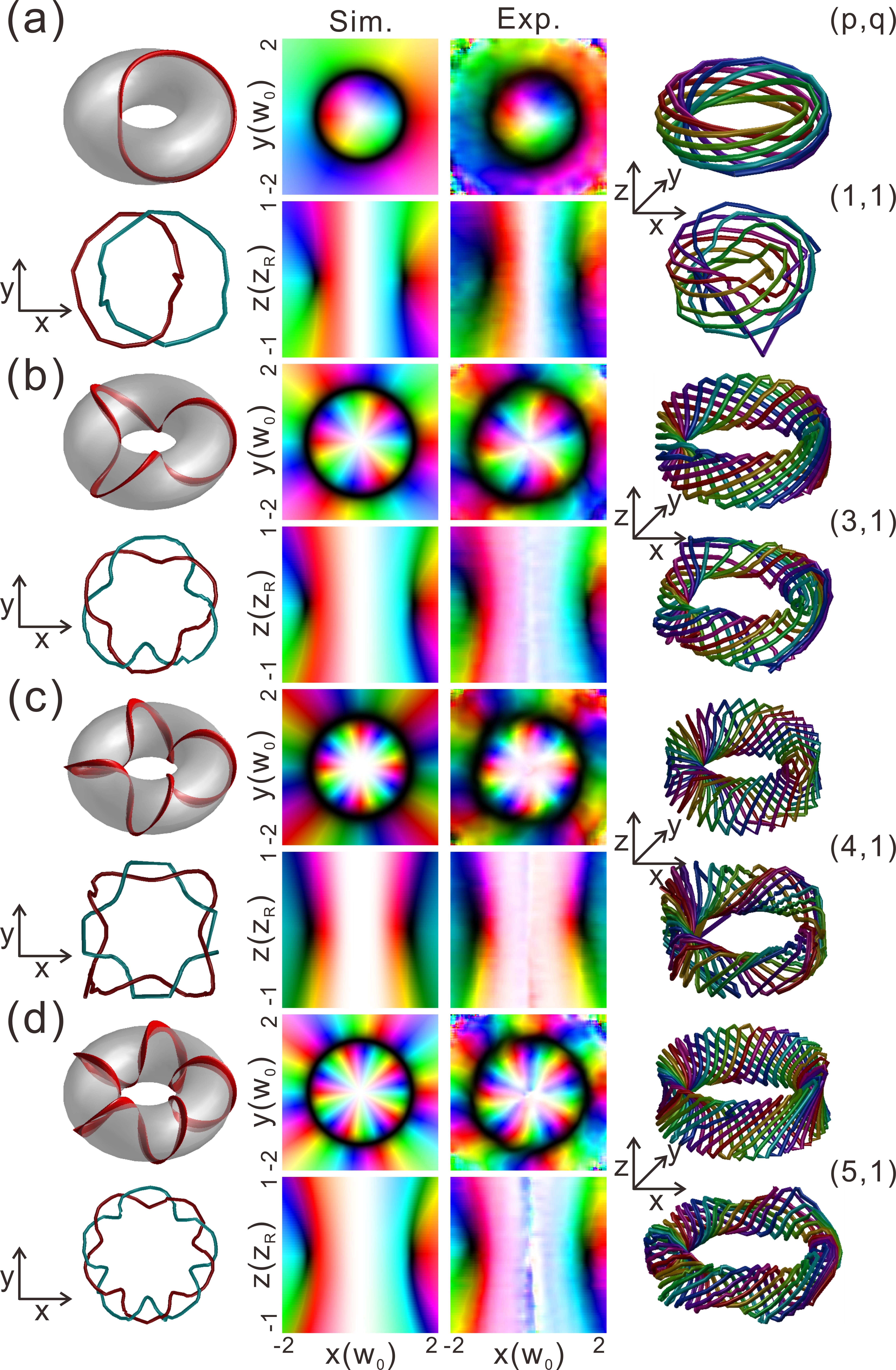}
        \\[0mm] 
        \caption{Simulation and experimental results of optical hopfions with winding number $(p,1)$. For (a), $p=1$; (b), $p=3$; (c), $p=4$ and (d), $p=5$. The middle column shows the polarization distributions in the $xy$ plane at $z=0$ and in the $xz$ plane at $y=0$; the upper left of each subplot is the torus knot corresponding to the respective order, and the lower left is a pair of orthogonal isopolar filaments with a fixed Stokes vector $\mathbf{S}$; on the right are the simulated and experimentally obtained Hopf fibration structures for $S_z = -0.5$. The corresponding coordinate axes for the 3D plot have been labeled.
} 
        \label{fig3}
    \end{figure*}

Constructing optical hopfions with arbitrary winding number $(p,1)$ hinges on the rational superposition of LG beams with different OAM at $z=0$, serving as the LCP and RCP components of the vector beam. In previous work \cite{shen2023topological}, the OAM of the RCP component was set to 0, while the OAM of the LCP component was increased. As the OAM of the LCP component increases, the effective overlap between the two orthogonal polarization modes decreases, making it impossible to form good hopfions within a finite volume. We assign adjacent OAM values to the two orthogonal modes, ensuring effective overlap regardless of the order $p$ \cite{zeng2025tailoring,wang2025generation}. The complex amplitude of the optical field at $z=0$ satisfies:
\begin{equation}\label{eq12}
    \mathbf{E}=\left \{ \begin{array}{rcl}
    \alpha\mathrm{LG}_0^{(p+1)/2}\mathbf{e_L}+[(1-\beta)\mathrm{LG}_1^{(-p+1)/2}+\beta\mathrm{LG}_0^{(-p+1)/2}]\mathbf{e_R}, & p\,\mbox{is odd}, \\
    \alpha\mathrm{LG}_0^{(p+2)/2}\mathbf{e_L}+[(1-\beta)\mathrm{LG}_1^{(-p+2)/2}+\beta\mathrm{LG}_0^{(-p+2)/2}]\mathbf{e_R}, & p\,\mbox{is even},
    \end{array}\right.
    \end{equation}
where $\alpha,\beta\in(0,1)$ are relative coefficients.

Fig.\ref{fig3} shows the simulated and experimental results for $ p = 1, 3, 4, 5, q=1$ obtained from the above equation. The middle column shows the polarization distributions of hopfions of different orders in the $xy$ plane at $z=0$ and in the $xz$ plane at $y=0$ for $\alpha=\beta=0.5$. Overall, the experimental results are consistent with the simulations. Since $q = 1$ here, a symmetrically distributed first-order skyrmion-anti-skyrmion pair structure can be observed in the $xz$ plane. For the $xy$ plane, a $p$-th order skyrmion exists within the region enclosed by the black (i.e., LCP) component. The left and right columns of Fig.\ref{fig3} show the Hopf fibration structures corresponding to the respective orders. The upper left of Fig.\ref{fig3} (a) shows the torus knot drawn according to Eq.\eqref{eq9}, which winds once in both the toroidal and poloidal directions, forming a circular shape. The lower part shows the experimentally reconstructed orthogonal isopolar filaments with $\mathbf{S} = (0.87, 0, -0.5)$ and $(-0.87, 0, -0.5)$, which are visibly linked once. The right side shows the simulated and experimentally reconstructed eight filaments with $S_z = -0.5$, forming a hollow torus structure. Since hopfions are generated within a finite volume, filaments with $S_z$ greater than a certain value are not closed, while those with $S_z$ less than this value form circular shapes. Figures (b), (c) and (d) use the same parameters as (a). As the order increases, the filaments exhibit a $p$-petal flower shape. However, all filaments with $S_z = -0.5$ still form a hollow torus. The difference is that the same color repeats $p$ times in the azimuthal direction.

In addition to the shape of the filaments, another tool for identifying the order of the generated hopfions is the Hopf index. According to Eq.\eqref{eq10}, the Hopf index is not only related to the quality of the generated Stokes vector but also to factors such as the volume of the integration region and the number of effective sampling points in the integration. Therefore, it is reasonable to compare the experimental results with the simulated integration results under the same parameter conditions, rather than directly comparing with the strict integer obtained from the product of the winding numbers. The experimentally obtained Hopf indices are $N_H = 0.6377, 2.1649, 2.5015, 3.1618$ for hopfions with $p = 1, 3, 4, 5,q = 1$, while the corresponding simulated results are $0.7519, 2.0758, 2.1101, 3.2436$. Due to the limited number of integration sampling points and the choice of $\alpha,\beta$ that results in a small effective integration region, the discrepancy between the integration result and the ideal value increases with the order. However, the expected results were achieved in terms of the shape of the filaments and the polarization distributions.

\subsection*{Construction of (1,q)-order optical hopfions}

\begin{figure*}[ht!]
    \vspace*{0mm} 
        \centering
        \includegraphics[height=8cm]{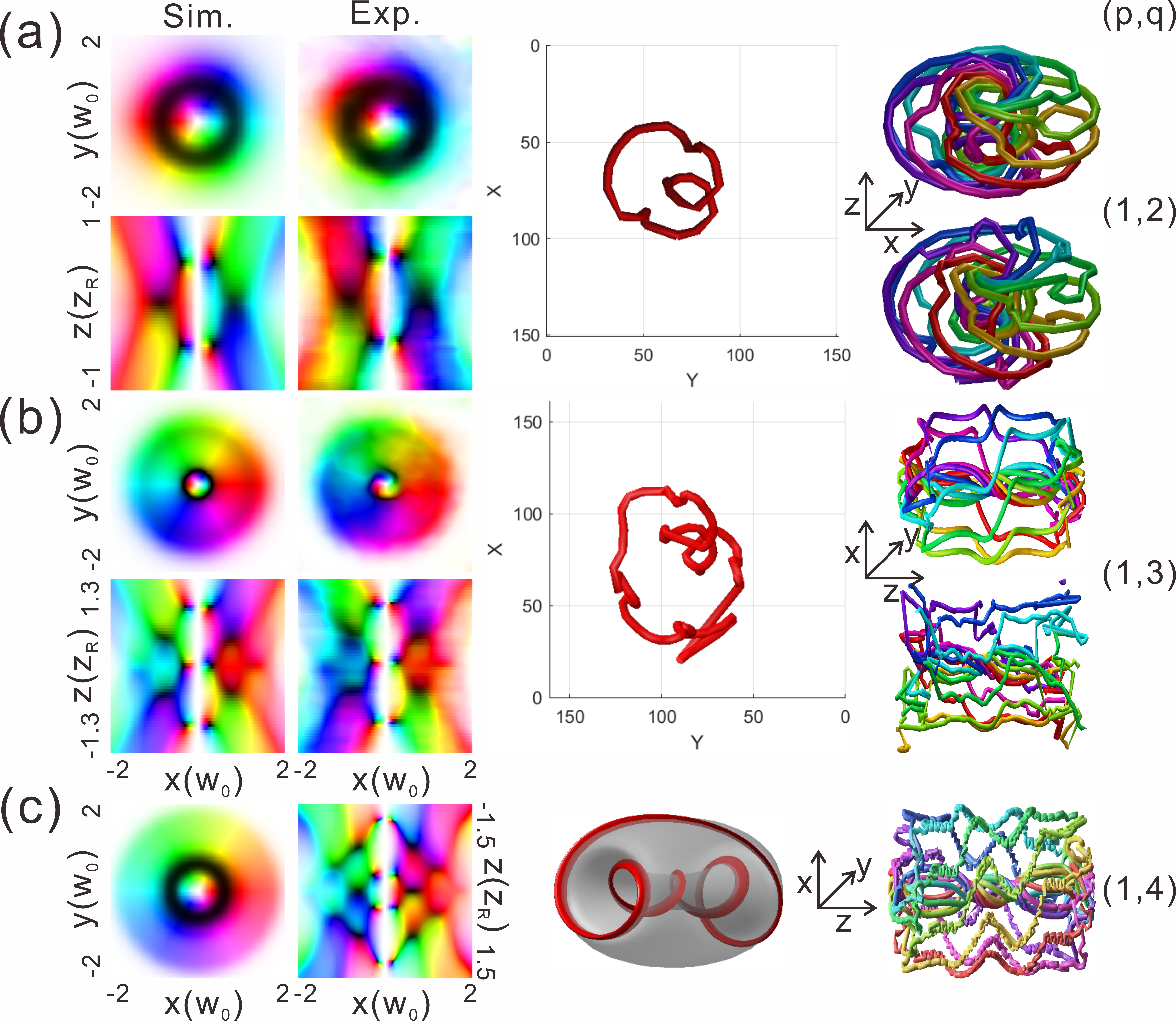}
        \\[0mm] 
        \caption{(a) and (b) are simulated and experimental results for $(1,q)$ winding hopfions. The left side show the polarization distributions in the $xy$ plane at $z=0$ and in the $xz$ plane at $y=0$; the middle show individual isopolar filaments, with coordinates representing sampling points; the right side show all filaments for a fixed $S_z$, with the upper part being the simulation and the lower part the experimental results. (c) shows the simulation results for $(1,4)$ hopfion, with the middle replaced by a torus knot drawn using parametric equations.
} 
        \label{fig4}
    \end{figure*}
Since there is no direct definition of OAM in the longitudinal ($xz$) plane, constructing hopfions with arbitrary $q$ remains highly challenging. However, considering that it essentially involves making equal polarizations encircle the z-axis more times during propagation, reasonably constructing the Gouy phase difference between the LCP and RCP components is the key to breakthrough. Thus, achieving arbitrary toroidal winding number $q$ requires careful control of the Gouy phase evolution through tailored superpositions of LG modes, enabling the polarization filaments to wind $q$ times around the z-axis. For a hopfion with winding number $(1, q)$, its complex amplitude should take the following form: 
\begin{equation}\label{eq13}
    \mathbf{E}=\alpha\mathrm{LG}_0^1\mathbf{e_L}+[(1-\beta)\sum_{i=1}^q(-1)^i\mathrm{LG}_i^0+\beta\mathrm{LG}_0^0]\mathbf{e_R},
\end{equation}
where $\alpha,\beta\in(0,1)$.    

Fig.\ref{fig4} (a), (b) show the simulation and experimental results of hopfions with winding numbers $p = 1$, $q = 2, 3 $ for $\alpha=0.5$, $\beta=0.4$. The left column of Fig.\ref{fig4} (a) shows the polarization distributions in the $xy$ plane at $z=0$ and in the $xz$ plane at $y=0$. Overall, the experimental results are consistent with the simulations. Since $p = 1$, there is only one first-order skyrmion structure in the $xy$ plane, the same color repeats only once. According to Eq.\eqref{eq8}, for $q = 2$ hopfions, a symmetric second-order skyrmion-anti-skyrmion pair structure should exist in the $xz$ plane. This is feasible with appropriate values of $\alpha$ and $\beta$. However, as seen in Fig.\ref{fig4} (a), under the chosen parameters, the second-order skyrmion in the $xz$ plane actually splits into two first-order skyrmions (just like the hopfions formed by tightly focused Zernike modes \cite{sugic2019unravelling}). Thanks to the topological stability of hopfions, the existence of this structure is also reasonable. To avoid visual confusion, only the experimental isopolar filament with $\mathbf{S} = (0.94, 0, -0.34)$ is shown in the middle of (a), which encircles the coordinate center twice: once as a small inner circle and once as a larger outer circle. The eight fibers with $S_z = -0.34$ on the right form a gyroid shape, with an outer layer of filaments and an inner core twisted into a single strand. For the case of $p=1, q=3$ in (b), the $xz$ plane contains three pairs of skyrmion and anti-skyrmion structure. The isopolar filaments encircle the center three times in total, including two small inner loops and one larger outer loop. All fibers satisfying $S_z = 0$ form an hourglass structure, with an outer layer of filaments and two inner strands twisted together. Note that the coordinate axis directions have been changed here due to the image size. Since the LG modes we used are normalized by $E_0$, the pattern should be maintained for higher-order $q$ values. That is, in the $xz$ plane, the polarization distributions consist of $q$ pairs of skyrmions and anti-skyrmions. The isopolar filaments encircle the z axis once on the outer side away from the z axis and $q-1$ times on the inner side. Fig.\ref{fig4} (c) shows the simulation results for $p=1$, $q=4$, with the torus knot clearly illustrating such winding structure. For higher-order $q$ values ($q>4$), the winding radius of the filaments around the center decreases, and the $S_z$ required to form a single closed filament increases.

For hopfions with $p=1$, $q=2,3$, the experimentally obtained Hopf indices are $N_H = 1.4107, 2.0786$, while the simulated results under the same parameters are $N_H = 1.7762, 2.5335$. Due to the small size of the skyrmions in the $xz$ plane, the experimental errors are amplified, but overall they are consistent with the simulations.
\subsection*{Combined (p,q)-order optical hopfions}

\begin{figure*}[ht!]
    \vspace*{0mm} 
        \centering
        \includegraphics[height=6cm]{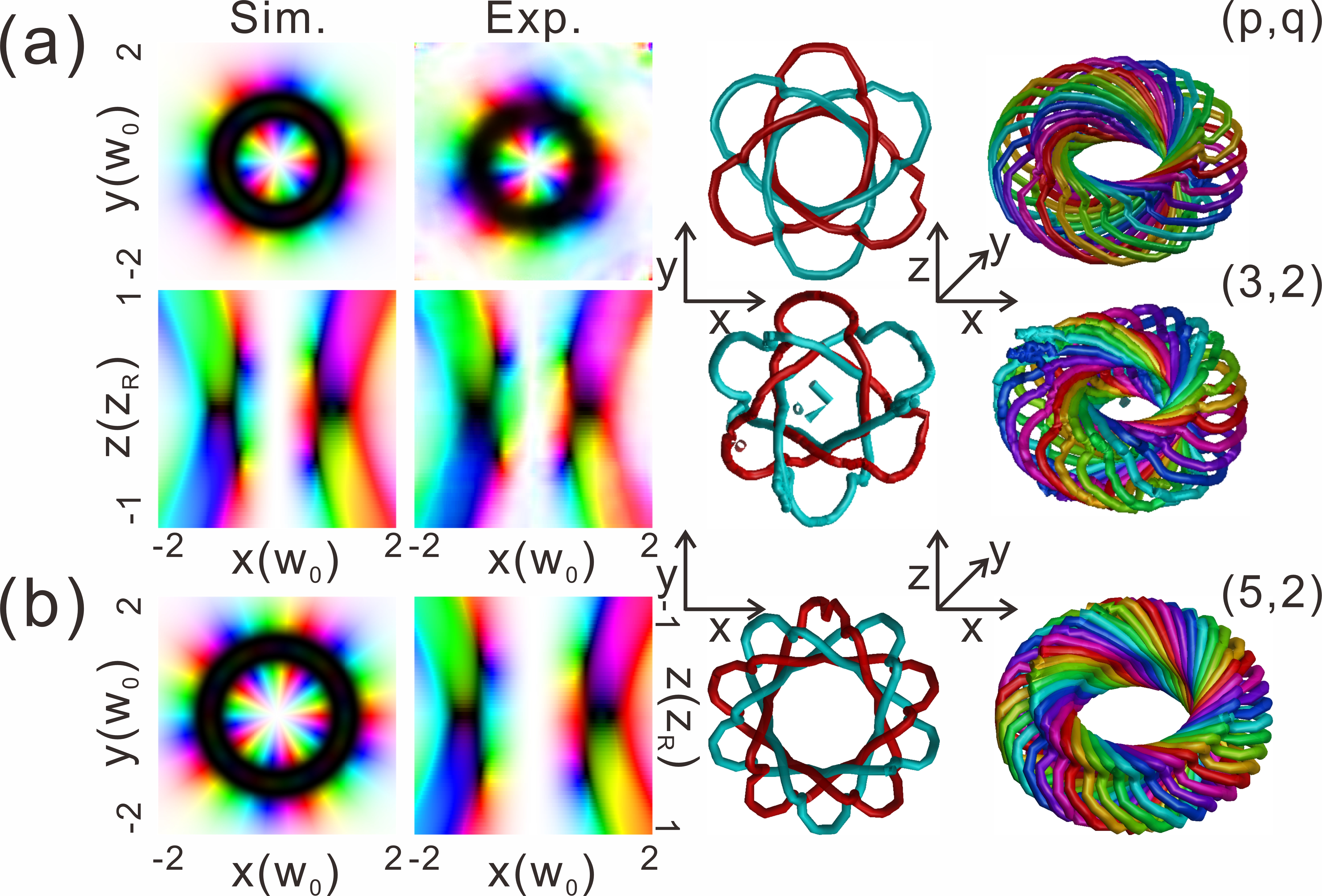}
        \\[0mm] 
        \caption{(a) shows the simulation and experimental results for the hopfion with winding number $(p,q) = (3,2)$, and (b) shows the simulation results for $(p,q) = (5,2)$. From left to right, they display the polarization distributions in the $xy$ and $xz$ planes, the mutually orthogonal isopolar filaments with $\mathbf{S} = (0.94, 0, -0.34)$ and $(-0.94, 0, -0.34)$, and the eight isopolar fibers with $S_z = -0.34$.
} 
        \label{fig5}
    \end{figure*}

Combining the two aforementioned parts can realize the generation of optical hopfions of arbitrary \((p,q)\) order, with the following form:
\begin{equation}\label{eq14}
    \mathbf{E}=\left \{ \begin{array}{rcl}
    \alpha\mathrm{LG}_0^{(p+1)/2}\mathbf{e_L}+[(1-\beta)\sum_{i=1}^q(-1)^i\mathrm{LG}_i^{(-p+1)/2}+\beta\mathrm{LG}_0^{(-p+1)/2}]\mathbf{e_R}, & p\,\mbox{is odd}, \\
    \alpha\mathrm{LG}_0^{(p+2)/2}\mathbf{e_L}+[(1-\beta)\sum_{i=1}^q(-1)^i\mathrm{LG}_i^{(-p+2)/2}+\beta\mathrm{LG}_0^{(-p+2)/2}]\mathbf{e_R}, & p\,\mbox{is even},
    \end{array}\right.
\end{equation}
where $\alpha,\beta\in(0,1)$.

 Fig.\ref{fig4} (a) shows the simulation and experimental results based on Eq.\eqref{eq14} for $(p,q) = (3,2)$, $\alpha=0.5$, $\beta=0.4$. Consistent with the previous patterns, there is a third-order skyrmion in the $xy$ plane at $z=0$, and two skyrmion-anti-skyrmion pairs structure in the $xz$ plane at $y=0$. Under these parameters, the separation between the two skyrmions along the z direction is relatively small. The isopolar filaments with $\mathbf{S} = (0.94, 0, -0.34)$ and $(-0.94, 0, -0.34)$ form the well-known trefoil knot shape, with the filaments crossing at the center to form a triangle and a three-petal flower structure on the periphery. For filaments of order $(p,q)$, the center should be a $p$-sided polygon, with $q-1$ layers of mutually orthogonal $p$-petal flower shapes extending radially from the center. The simulation results for $(p,q) = (5,2)$ under the same parameters, shown in Fig.\ref{fig5} (b), precisely demonstrate this structure. The right side of Fig.\ref{fig5} show the eight filament structures for $S_z = -0.34$, with the experimental structures being essentially consistent with the simulations. The Hopf index calculated for the $(3,2)$ order hopfion obtained experimentally is 5.2751, corresponding to the simulated result of 5.3950. It should be noted that to maintain a singly connected shape of the filaments, $p$ and $q$ must be coprime. If this condition is not met, the filaments will split into $q$ parts (assuming $p \geq q$), each with a winding number of $(p/q, 1)$ (see supplementary).
\subsection*{The impact of coefficient values}

\begin{figure*}[ht!]
    \vspace*{0mm} 
        \centering
        \includegraphics[height=5cm]{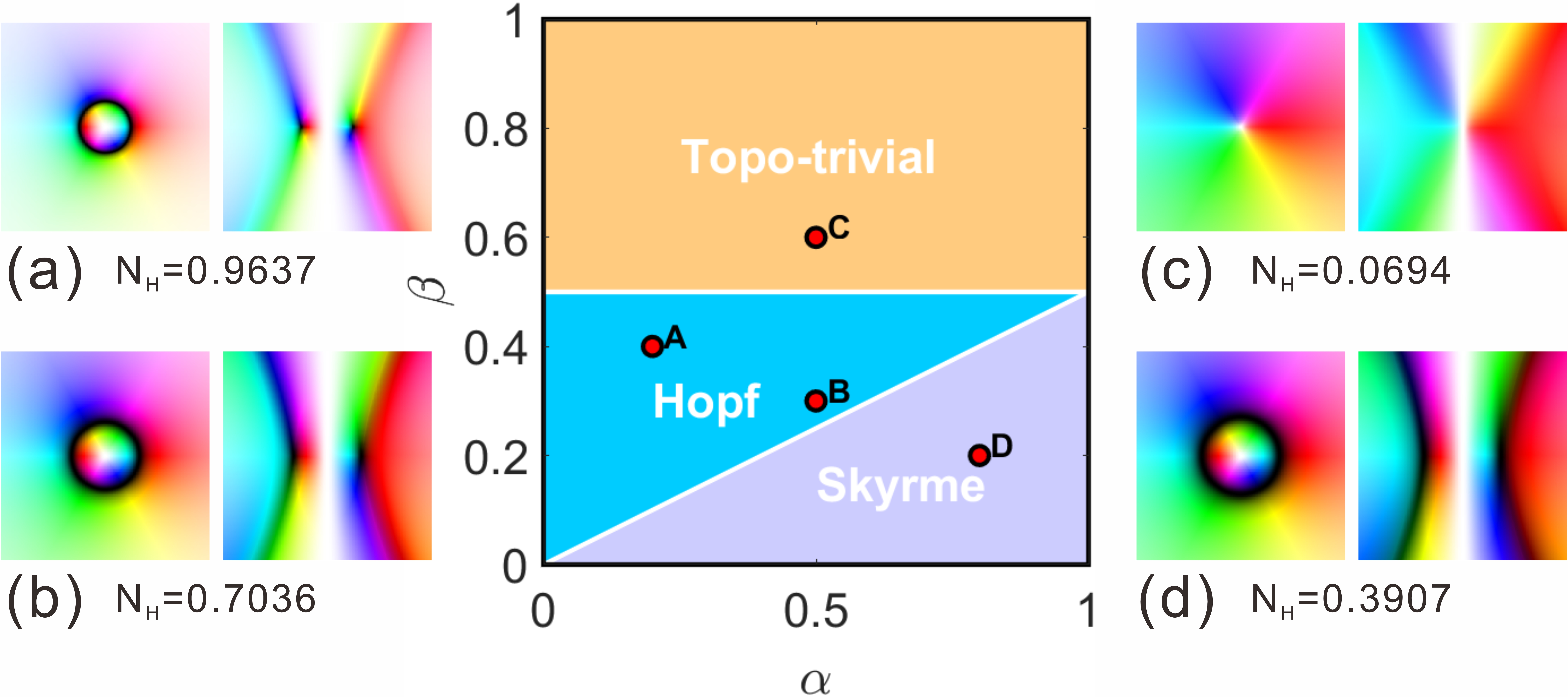}
        \\[0mm] 
        \caption{Polarization distributions in the xy and xz planes for different $\alpha$ and $\beta$ values. The middle shows three regions: hopfion, skyrmion tube, and topologically trivial. Points A, B, C, and D are given by (a), (b), (c), and (d), respectively.
} 
        \label{fig6}
    \end{figure*}

The reliability of the generated optical hopfion structure is highly sensitive to the relative weights $\alpha$ and $\beta$ in the superposition. In this section, we systematically explore their impact. According to Eq.\eqref{eq14}, Fig.\ref{fig6} shows the polarization distributions in the $xy$ and $xz$ planes for four sets of $(\alpha, \beta)$ values—$A(0.2, 0.4)$, $B(0.5, 0.3)$, $C(0.5, 0.6)$, and $D(0.8, 0.2)$—for hopfions with $(p,q) = (1,1)$ within the range $\alpha,\beta\in(0,1)$. The hopfions region is located at $0 < \alpha < 1$, $\alpha/2 < \beta < 1/2$. Points A and B are two examples within this region, both showing good hopfion structures. However, due to the different numbers of polarization states contained within the finite volume, the calculated Hopf indices will vary. Point C is located in the topological-trivial region. Its topological structure is completely destroyed as seen from the polarization distributions, and the calculated Hopf index is very low. Point D is in a special region. Although it can be considered a hopfion in an infinitely large space, it forms a skyrmion tube within a finite volume. The isopolar filaments cannot close, and thus the calculated Hopf index is also low. For higher-order $p$ values, the corresponding hopfion regions have different $\alpha$ and $\beta$ values, but as long as they are within the appropriate range, well-defined hopfion structures can be obtained. It should be noted that such profiles differ for different orders; therefore, appropriate parameters must be selected to construct more desirable optical hopfions.

\section*{Discussion}
In this work, we present a general method to construct optical hopfions of arbitrary winding number $(p,q)$ by coherently superposing LG modes with carefully designed Gouy phase. Via the Hopf mapping, we obtain the explicit analytical expressions for polarization hopfions of arbitrary order, providing direct guidance for the target beam design. Experimentally, these hopfions are generated and fully characterized in 3D free space using complex amplitude modulated SLM combined with digital propagation. We also provide the analytical expressions for $(p,1)$ and $(1,q)$ configurations and extend to the general $(p,q)$ case through superposition. Several examples are experimentally realized, with the reconstructed polarization fields and isopolarization filaments confirming the design, including previously unreported arbitrary-order states with toroidal winding number $q$.

The design principle, though implemented optically, is transferable to other physical systems hosting vector-field topological excitations. The formal analogy between the optical Stokes vector and fields such as magnetization in chiral magnets or the director in liquid crystals suggests that a similar superposition strategy, using appropriately tailored excitations, could nucleate arbitrary-order hopfions in those platforms. Thus, our work bridges structured light and topological matter, providing a conceptual blueprint for synthesizing complex 3D topological textures across disciplines.

Looking forward, arbitrary-order hopfions offer promising opportunities as robust, high-dimensional information carriers in photonics and beyond. Each distinct $(p,q)$ pair could encode a topological state for information processing. Immediate next steps include exploring their cross-band conversion \cite{wu2022conformal}, non-local quantum generation \cite{ornelas2024non}, stability in complex media \cite{wang2024topological}, and integration into photonic devices \cite{luo2021skyrmion}. Further refinement of generation reliability, potentially via adaptive optics \cite{hampson2021adaptive}, will be valuable for applications that require precise control of topological charge. This platform, by bridging versatile beam shaping with topological concepts, opens a potential avenue for investigating and utilizing 3D topological structures in both fundamental and applied physics.

\section*{Methods}
\subsection*{Numerical calculation of the Hopf index}
According to \cite{liu2018binding}, we present here the detailed process for numerically calculating the Hopf index $N_H$ when $\mathbf{S}(\mathbf{x}) = (S_x, S_y, S_z)$ is known, where $\mathbf{x}=(x,y,z)$. Based on the pitch form of $\mathbf{B}(\mathbf{x})$ (note the coefficient $1/2$ is not include), the explicit expressions for its three components can be obtained:
\begin{equation}\label{eq15}
\begin{split}
    B_x&=\mathbf{S}(\mathbf{x})\cdot(\frac{\partial\mathbf{S}(\mathbf{x})}{\partial y}\times\frac{\partial\mathbf{S}(\mathbf{x})}{\partial z}),\\
    B_y&=\mathbf{S}(\mathbf{x})\cdot(\frac{\partial\mathbf{S}(\mathbf{x})}{\partial z}\times\frac{\partial\mathbf{S}(\mathbf{x})}{\partial x}),\\
    B_z&=\mathbf{S}(\mathbf{x})\cdot(\frac{\partial\mathbf{S}(\mathbf{x})}{\partial x}\times\frac{\partial\mathbf{S}(\mathbf{x})}{\partial y}),
    \end{split}
\end{equation}
To avoid solving the explicit form of $\mathbf{A}(\mathbf{x})$, we perform Fourier transforms on $\mathbf{A}(\mathbf{x})$ and $\mathbf{B}(\mathbf{x})$ to obtain $\mathbf{A}(\mathbf{k}) = \mathcal{F}(\mathbf{A}(\mathbf{x}))$ and $\mathbf{B}(\mathbf{k}) = \mathcal{F}(\mathbf{B}(\mathbf{x}))$, where $\mathbf{k} = (k_x, k_y, k_z)$ is the momentum space coordinate. Then, $\nabla \times \mathbf{A}(\mathbf{x}) = \mathbf{B}(\mathbf{x})$ becomes:
\begin{equation}\label{eq16}
    i2\pi\mathbf{k}\times\mathbf{A}(\mathbf{k})=\mathbf{B}(\mathbf{k}),
\end{equation}
using the Lagrange formula and the gauge relation $\mathbf{k} \cdot \mathbf{A}(\mathbf{k}) = 0$, we can solve for:
\begin{equation}\label{eq17}
    \mathbf{A}(\mathbf{k})=-i\frac{\mathbf{k}\times\mathbf{B}(\mathbf{k})}{2\pi\mathbf{k}^2},
\end{equation}
then, by using the inverse Fourier transform, we obtain $\mathbf{A}(\mathbf{x}) = \mathcal{F}^{-1}(\mathbf{A}(\mathbf{k}))$. Assuming the number of sampling points in the three directions are $N_x$, $N_y$, and $N_z$, respectively, we finally obtain the numerical form of the Hopf index:
\begin{equation}\label{eq18}
    N_H=\frac{1}{(4\pi)^2}\frac{1}{N_xN_yN_z}\sum_{\mathbf{x}}\mathbf{B}(\mathbf{x})\cdot\mathbf{A}(\mathbf{x}),
\end{equation}
note that if the differential and integral intervals are taken as unit 1, the corresponding unit momentum space coordinates need to be determined, and Eq.\eqref{eq18} becomes $N_H = 1/(4\pi)^2 \sum_{N_x, N_y, N_z} \mathbf{B}(\mathbf{x}) \cdot \mathbf{A}(\mathbf{x})$.

\subsection*{Angular spectrum method for digital propagation}
The complex amplitude distribution $E(x,y)$ of a monochromatic light field in a plane can be regarded as the superposition of monochromatic plane waves propagating in different directions, with the weight factors $A(f_x,f_y)$ of each direction being the angular spectrum of $E(x,y)$, where $A(f_x,f_y)=\mathcal{F}(E(x,y))$, $f_x,f_y$ are the spatial frequency coordinates. The angular spectrum also satisfies the Helmholtz equation, but only retains differentiation in the z-direction ($(\nabla_z^2+k_z^2)A=0$):
\begin{equation}\label{eq19}
\frac{\mathrm{d}^2A(f_x,f_y)}{\mathrm{d}z^2}+k^2[1-\lambda^2(f_x^2+f_y^2)]A(f_x,f_y)=0,    
\end{equation}
the solution of Eq.\eqref{eq19} is $A(f_x,f_y)=A_0(f_x,f_y)\exp{[ikz\sqrt{1-\lambda^2(f_x^2+f_y^2)}]}$, where $A_0(f_x,f_y)=\mathcal{F}(E_0(x,y))$, $E_0(x,y)$ is the complex amplitude distribution in $z=z_0=0$. Thus, once we know the initial complex amplitude distribution, we can obtain the angular spectrum $A(f_x, f_y)$ at the propagation distance $z$. Performing the inverse Fourier transform $E(x,y) = \mathcal{F}^{-1}(A(f_x, f_y))$ yields the complex amplitude distribution to be loaded onto the SLM. Such digital propagation technique does not require lenses for Fourier transformation, thereby reducing experimental errors to some extent.


\section*{Acknowledgements}

National Natural Science Foundation of China (No.92476105, 12404390, 12304406, 92050103 and 12104358); Shaanxi Fundamental Science Research Project for Mathematics and Physics (No. 22JSQ035); Shaanxi Province postdoctoral Science Foundation (No.2023BSHEDZZ23). Singapore Ministry of Education (MOE) (RG157/23 \& RT11/23), Singapore
Agency for Science, Technology and Research (A*STAR) (M24N7c0080 \& 25619013), and
Nanyang Assistant Professorship Start Up grant.

\section*{Author contributions statement}
Y.S., A.F. and H.G. conceived the project. X.Z. and J.W., and G.L. performed the experiments. X.Z., Y.C., Z.G., Y.Z., and X.Y. analyzed the data. X.Z., J.W., Y.C., G.L., D.W., H.C., X.Y., C.W., Y.S., A.F. and H.G. contributed to the experimental setup and interpretation of the results. Y.S., A.F. and H.G. supervised the project. All authors contributed to writing the manuscript.

\section*{Additional information}
\textbf{Competing financial interests:} The authors declare no competing financial interests.

\end{document}


\renewcommand{\figurename}{Fig. S}

\title{Supplemental Material: Optical hopfions with arbitrary two winding numbers}

\author{Xinji Zeng}\email{These authors contributed equally to this work.}\author{Jinwen Wang}\email{These authors contributed equally to this work.}
\affiliation{Ministry of Education Key Laboratory for Nonequilibrium Synthesis and Modulation of Condensed Matter, Shaanxi Province Key Laboratory of Quantum Information and Quantum Optoelectronic Devices, School of Physics, Xi'an Jiaotong University, Xi'an 710049, China}
\author{Yun Chen}\email{These authors contributed equally to this work.}
\affiliation{Department of Physics, Huzhou University, Huzhou 313000, China}
\author{Guang Liu}
\affiliation{Ministry of Education Key Laboratory for Nonequilibrium Synthesis and Modulation of Condensed Matter, Shaanxi Province Key Laboratory of Quantum Information and Quantum Optoelectronic Devices, School of Physics, Xi'an Jiaotong University, Xi'an 710049, China}
\author{Zhenyu Guo}
\affiliation{Center for Disruptive Photonic Technologies, School of Physical and Mathematical Sciences $\And$ The Photonics Institute, Nanyang Technological University, Singapore 637371, Singapore}
\author{Yongkun Zhou}\author{Xin Yang}\author{Chengyuan Wang}\author{Haixia Chen}\author{Dong Wei}
\affiliation{Ministry of Education Key Laboratory for Nonequilibrium Synthesis and Modulation of Condensed Matter, Shaanxi Province Key Laboratory of Quantum Information and Quantum Optoelectronic Devices, School of Physics, Xi'an Jiaotong University, Xi'an 710049, China}
\author{Yijie Shen}
\email{Yijie.shen@ntu.edu.sg}
\affiliation{Center for Disruptive Photonic Technologies, School of Physical and Mathematical Sciences $\And$ The Photonics Institute, Nanyang Technological University, Singapore 637371, Singapore}
\affiliation{School of Electrical and Electronic Engineering, Nanyang Technological University, Singapore, 639798, Singapore}
\author{Andrew Forbes}
\email{andrew.forbes@wits.ac.za}
\affiliation{School of Physics, University of the Witwatersrand, Johannesburg, South Africa.}
\author{Hong Gao}
\email{honggao@xjtu.edu.cn}
\affiliation{Ministry of Education Key Laboratory for Nonequilibrium Synthesis and Modulation of Condensed Matter, Shaanxi Province Key Laboratory of Quantum Information and Quantum Optoelectronic Devices, School of Physics, Xi'an Jiaotong University, Xi'an 710049, China}
\affiliation{State Key Laboratory of Human-Machine Hybrid Augmented Intelligence, Xi’an Jiaotong University, Xi’an 710049,
China}

\maketitle
Here we present the mathematical formulation of hopfions together with supplementary experimental and numerical details. This includes: (1), Topological derivation of hopfions; (2), Experimental setup; (3), Stokes vector distribution of a typical $(p,q)$-order hopfion; (4), Hopfions with non-coprime winding numbers.

\section{Topological derivation of hopfions}

 Both skyrmions and hopfions originate essentially from stereographic projection in their mapping from parameter space to real space. Therefore, we first introduce this concept. We consider the one-point compactification of $\mathbb{R}^n$ to $S^n$, where the north pole of $ S^n$ is tangent to the origin of $\mathbb{R}^n$, and the projection point is the south pole $(0, \ldots, 0, -1)$, i.e., the following topological equivalence:
\begin{equation}\label{eq1}
    \mathbb{R}^n\cup \{\infty\}\cong S^n\subset \mathbb{R}^{n+1},
\end{equation}
which can be achieved by the stereographic projection:
\begin{equation}\label{eq2}
    f:\mathbb{R}^n\rightarrow S^n,\mathbf{x}\mapsto(\frac{2\mathbf{x}}{1+\|\mathbf{x}\|^2},\frac{1-\|\mathbf{x}\|^2}{1+\|\mathbf{x}\|^2}),
\end{equation}
where $\mathbf{x}=(x_1,...,x_n)\in\mathbb{R}^n$, $f(\mathbf{x})=(X_1,...,X_{n+1})\in S^n$ and satisfies $\sum_{i=1}^{n+1} X_i^2=1$. The Stokes vector $\mathbf{S}=(S_x,S_y,S_z)$, representing the polarization distribution of a light beam, is defined in the 3D Euclidean space $\mathbb{R}^3$. According to Eq.\eqref{eq2}, $\mathbb{R}^3$ can be compactified to $S^3$:
\begin{equation}\label{eq3}
    g:\mathbf{r}=(x,y,z)\mapsto(X_1,X_2,X_3,X_4)=(\frac{2x}{1+r^2},\frac{2y}{1+r^2},\frac{2z}{1+r^2},\frac{1-r^2}{1+r^2}).
\end{equation}
View $S^3$ as the unit hypersphere in $\mathbb{C}^2$, i.e., $S^3 = \{(Z_1, Z_2) \in \mathbb{C}^2 : |Z_1|^2 + |Z_2|^2 = 1\}$, where $Z = (Z_1, Z_2)^T = (X_4 + iX_3, X_1 + iX_2)^T$ is the spinor representation in the $S^3$ coordinate system. The normalized Stokes vector $\mathbf{S}$ represents the 2D sphere $S^2$, also known as the Poincaré sphere, and thus the Hopf map can be defined \cite{wilczek1983linking}:
\begin{equation}\label{eq4}
    \pi:\mathbb{C}^2\rightarrow S^2,(Z_1,Z_2)\mapsto \mathbf{S}=(S_x,S_y,S_z)=Z^{\dagger}\mathbf{\sigma}Z,
\end{equation}
where $\mathbf{\sigma}=(\sigma_x,\sigma_y,\sigma_z)$ are the Pauli matrices, expanding the above expression yields $(S_x, S_y, S_z) = (2\mathrm{Re}(Z_2Z_1^*), 2\mathrm{Im}(Z_2Z_1^*),\\ |Z_1|^2 - |Z_2|^2)$. Thus, according to Eq.\eqref{eq3}, we have obtained the expression for $\mathbf{S}$ representing the hopfion defined in $\mathbb{R}^3$. However, to more clearly reveal the topological structure and introduce higher-dimensional degrees of freedom, we introduce a complex function $w$ and consider the projection from $\mathbb{C}$ to $S^2$:
\begin{equation}\label{eq5}
    h:\mathbb{C}\rightarrow S^2,w=u+iv\mapsto\mathbf{S}=(\frac{2u}{1+|w|^2},\frac{2v}{1+|w|^2},\frac{1-|w|^2}{1+|w|^2}),
\end{equation}
the inverse mapping $h^{-1}$ gives $w = \frac{S_x + iS_y}{1 + S_z} = \frac{2Z_2Z_1^*}{2|Z_1|^2} = \frac{Z_2}{Z_1}=\frac{2(x+iy)}{1-r^2+2iz}$, where we have used the spinor representation and Eq.\eqref{eq3}. Introducing the toroidal coordinates $\mathbf{r}(\eta, \theta, \phi)$, which are related to the cylindrical coordinates $(\rho, \phi, z)$ as follows:
\begin{equation}\label{eq6}
    \rho=a\frac{\sinh{(\eta)}}{\cosh{(\eta)}-\cos{(\theta)}},\phi=\phi,z=a\frac{\sin{(\theta)}}{\cosh{(\eta)}-\cos{(\theta)}},
\end{equation}
where the toroidal parameter $\eta\in(0,\infty)$, the poloidal angle $\theta\in[0,2\pi]$, the toroidal angle $\phi\in[0,2\pi]$ and $a$ is a scal parameter. For a hopfion with toroidal winding number $q$ and poloidal winding number $p$, $w$ can be defined in the following form \cite{guslienko2023emergent}: 
\begin{equation}\label{eq7}
    w=\frac{(Z_2)^q}{(Z_1)^p}=\cosh^p{(\eta)}\tanh^q{(\eta)}e^{i(p\theta+q\phi)},
\end{equation}
substituting Eq.\eqref{eq7} into Eq.\eqref{eq5} yields the analytical expression for hopfions with $(p, q)$ winding numbers:
\begin{equation}\label{eq8}\begin{split}
    S_z&=P\frac{1-\cosh^{2p}{(\eta)}\tanh^{2q}{(\eta)}}{1+\cosh^{2p}{(\eta)}\tanh^{2q}{(\eta)}},\\
    S_x+iS_y&=\sqrt{1-S_z^2}\exp{[i(p\theta+q\phi)]},
    \end{split}
\end{equation}
where $P=\pm1$ is the polarity. For a given Stokes vector $\mathbf{S}=\mathbf{S}_0$ on a hopfion, the isoclines of polarization form a torus knot in 3D space, also described by the winding numbers $(p, q)$. For Eq.\eqref{eq6}, when $\eta = \eta_0$ is constant and $\theta = pt$, $\phi = qt$ with $t \in [0, 2\pi]$, the parametric equations for the torus knot can be obtained:
\begin{equation}\label{eq9}
    \begin{split}
        x(t)&=[R+r\frac{\cos{(pt)}}{R/r-\cos{(pt)}}]\cos{(qt)},
        \\y(t)&=[R+r\frac{\cos{(pt)}}{R/r-\cos{(pt)}}]\sin{(qt)},\\
        z(t)&=ar\frac{\sin{(pt)}}{R/r-\cos{(pt)}},
    \end{split}
\end{equation}
where we have used $x=\rho\cos{(\phi)},y=\rho\sin{(\phi)}$, the toroidal radius $R=\coth{(\eta_0)}$, the poloidal radius $r=1/\sinh{(\eta_0)}$. We have thus obtained the Stokes vector expression  Eq.\eqref{eq8} for optical hopfions in 3D space, with their polarization distribution satisfying the Hopf map from the hypersphere $S^3$ to the Poincar\'e sphere $S^2$. For a specific polarization state in 3D space, the fibers form a torus knot with winding numbers $(p, q)$, described by Eq.\eqref{eq9}.

\section{Experimental setup}
    
\begin{figure*}[ht!]
    \vspace*{0mm} 
        \centering
        \includegraphics[height=5cm]{FIGS1.jpg}
        \\[0mm] 
        \caption{Experimental setup. HWP. half-wave plate; PBS, polarization beam splitter; BD, beam displacer; BS, beam splitter; DHWP, D-shaped half-wave plate; SLM, spatial light modulator; L1,L2, lens; F, filter; QWP, quarter-wave plate; CCD, charge-coupled device.
} 
        \label{fig2}
    \end{figure*}
    
The experimental setup we adopted is shown in Fig.S\ref{fig2}. A laser with a wavelength of 780 nm generated by an external cavity semiconductor laser is shaped through a single-mode fiber and then incident on the experimental system with a good Gaussian mode. Before the beam enters the BD and is split, it first passes through a combination of a HWP and a PBS to adjust the incident light intensity, and then passes through a HWP with its fast axis oriented at 45 degrees to convert the horizontally polarized light into diagonally polarized. The BD splits the beam into two beams with polarizations oriented horizontally and vertically, respectively. After passing through a BS, the two beams are prepared to be incident on the SLM. Since the SLM only modulates horizontally polarized light, a DHWP is employed to convert the vertically polarized light into horizontally polarized. After being reflected by the SLM, the beam passes through the DHWP again and is converted back to vertically polarized, which facilitates the subsequent beam combination using the BD. The screen of the SLM is divided into two parts. Using the complex amplitude modulation technique, the left-hand and right-hand circularly polarized (LCP and RCP) orthogonal modes of the target vector beam, as well as the same blazed grating, are loaded respectively. The waist size of the target light field mode is set to 0.25 mm. The BS reflects the modulated beam into a 4f imaging system, where a filter at the second focal point selects the +1 diffraction orders of the two beams as the target beams. After passing through two lenses, the target beams are combined in another BD and converted into orthogonal circular polarizations by a QWP to form the target vector beam. 

The CCD is placed at the last focal point and does not need to be moved during the experiment. Instead, the target modes at different propagation distances $z$ are loaded onto the SLM using the angular spectrum method. A Stokes measurement setup in front of the CCD, consisting of an HWP, QWP, and PBS, is used to reconstruct the 3D polarization distribution of the target vector beam.

\section{Expression of the Laguerre–Gaussian mode}

Under the paraxial condition, beam-like solutions are described by the scalar Helmholtz equation $(\nabla^2 + k^2)E = 0$, where $k$ is the wavenumber, with a particular solution being the LG mode:
\begin{equation}\label{eq11}
    \mathrm{LG}^l_p(r,\phi,z)=\frac{E_0}{w(z)}(\frac{\sqrt{2}r}{w(z)})^{|l|}\exp(-\frac{r^2}{w^2(z)})\mathrm{L}^{|l|}_p(\frac{2r^2}{w^2(z)})\exp(il\phi)\exp(-i\frac{kr^2}{2R(z)})\exp(i\Phi(z)),
\end{equation}
where $E_0=\sqrt{\frac{2p!}{\pi(|l|+p)!}}$, $w(z)=w_0\sqrt{1+(z/z_R)^2}$, $w_0$ is the beam waist and $z_R=\pi w_0^2/\lambda$ is the Rayleigh length with $\lambda$ the wavelength, $R(z)=z(1+(z_R/z)^2)$ is the radius of curvature and $\Phi(z)=(2p+|l|+1)\arctan{(z/z_R)}$ is the Gouy phase, $\mathrm{L}^{|l|}_p(\cdot)$ is the associated Lagueree polynomials.

\section{Stokes vector distribution of a typical (p,q)-order hopfion}

\begin{figure*}[ht!]
    \vspace*{0mm} 
        \centering
        \includegraphics[height=22cm]{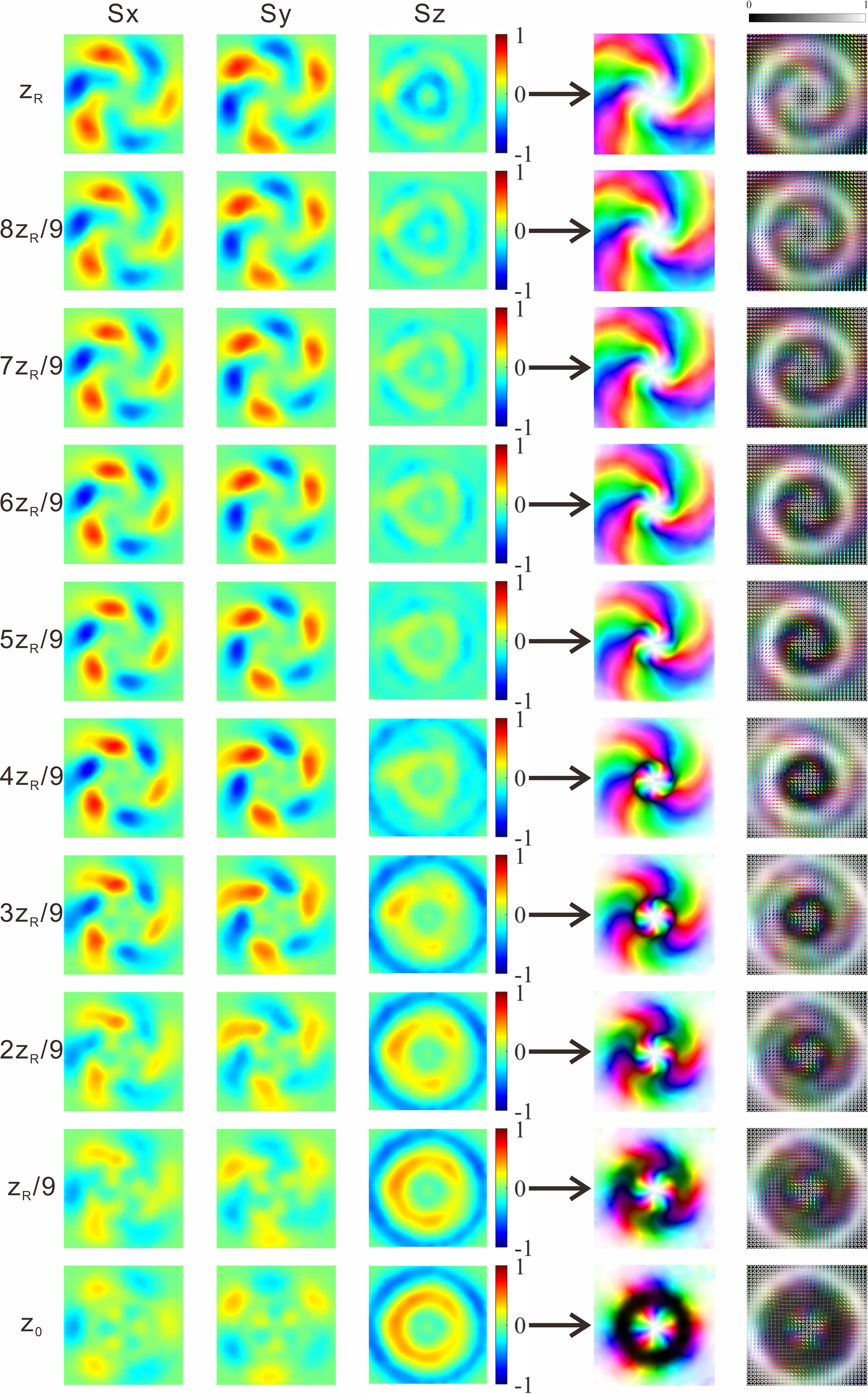}
        \\[0mm] 
        \caption{Stokes vectors and polarization distributions of a $(p,q)=(3,2)$ hopfion: nine axial slices from $z=0$ to $z=z_R$, polarization colormap  as in the main text, with grayscale light intensity shown on the right.
} 
        \label{figs2}
    \end{figure*}
    
To visualize how the LCP and RCP components of an asymmetric, non-intrinsic vector field (relative to cylindrical vector beams) evolve into a hopfion configuration during propagation, we present in Fig. S\ref{figs2} the transverse Stokes vectors and polarization distributions of a $(p,q)=(3,2)$ hopfion from $z=0$ to $z=z_R$.

The polarization maps use the same colormap as the main text. Overall, the key difference from cylindrical vector beams is that the polarization pattern evolves continuously during propagation, arising from the Gouy-phase imbalance between the asymmetric LCP and RCP components. The Stokes parameter $S_z$ in the figure encodes their relative weights: dark red denotes pure RCP, dark blue pure LCP. As the propagation distance increases, the distribution of $S_z$ reveals a gradual transition in which the LCP component supplants the RCP component as the dominant polarization. This evolution is also evident in the behavior of the black ring (denoting RCP) within the polarization maps: the initially thick black ring progressively weakens and is overlain by other colors, briefly re-emerges as a narrow ring, and is ultimately submerged once again. If we isolate and track the evolution of a single polarization lobe, the mechanism by which the field twists into a torus knot in 3D space becomes transparent. Take one of the three azimuthal green petals as an example: the outer green segment first crosses the black ring and merges with its nearest inner counterpart; immediately afterward the two are split by a thin black ring, while the entire inner set rotates counter-clockwise. When the inner green lobe approaches the next outer green segment, the two reconnect. This sequence is the direct embodiment of a winding number $q = 2$ around the torus axis: identical polarization filaments execute a full $2\pi$ rotation about the propagation axis before rejoining their outer partners, thereby completing two closed circuits and producing the observed double winding.

The polarization and intensity maps on the right-hand side furnish additional insight. The skyrmion embedded in the transverse structure of the hopfion exhibits a star-shaped polarization texture, a signature that can be exploited for a finer topological classification of hopfions; throughout propagation this texture merely undergoes a gentle twist, leaving its essential winding invariant. The intensity is rendered as a grayscale backdrop. Because the LCP and RCP components carry orbital angular momenta $l = –1$ and $l = 2$, respectively, the overall fluence naturally adopts a hollow ring profile. Moreover, the LCP beam itself is a coherent superposition of LG modes with radial indices $p = 0, 1, 2$; as propagation proceeds, the dominant ring of this composite intensity migrates radially, thereby sculpting in 3D the prescribed polarization skeleton of the targeted hopfion.

\section{Hopfions with non-coprime winding numbers}
\begin{figure*}[ht!]
    \vspace*{0mm} 
        \centering
        \includegraphics[height=9cm]{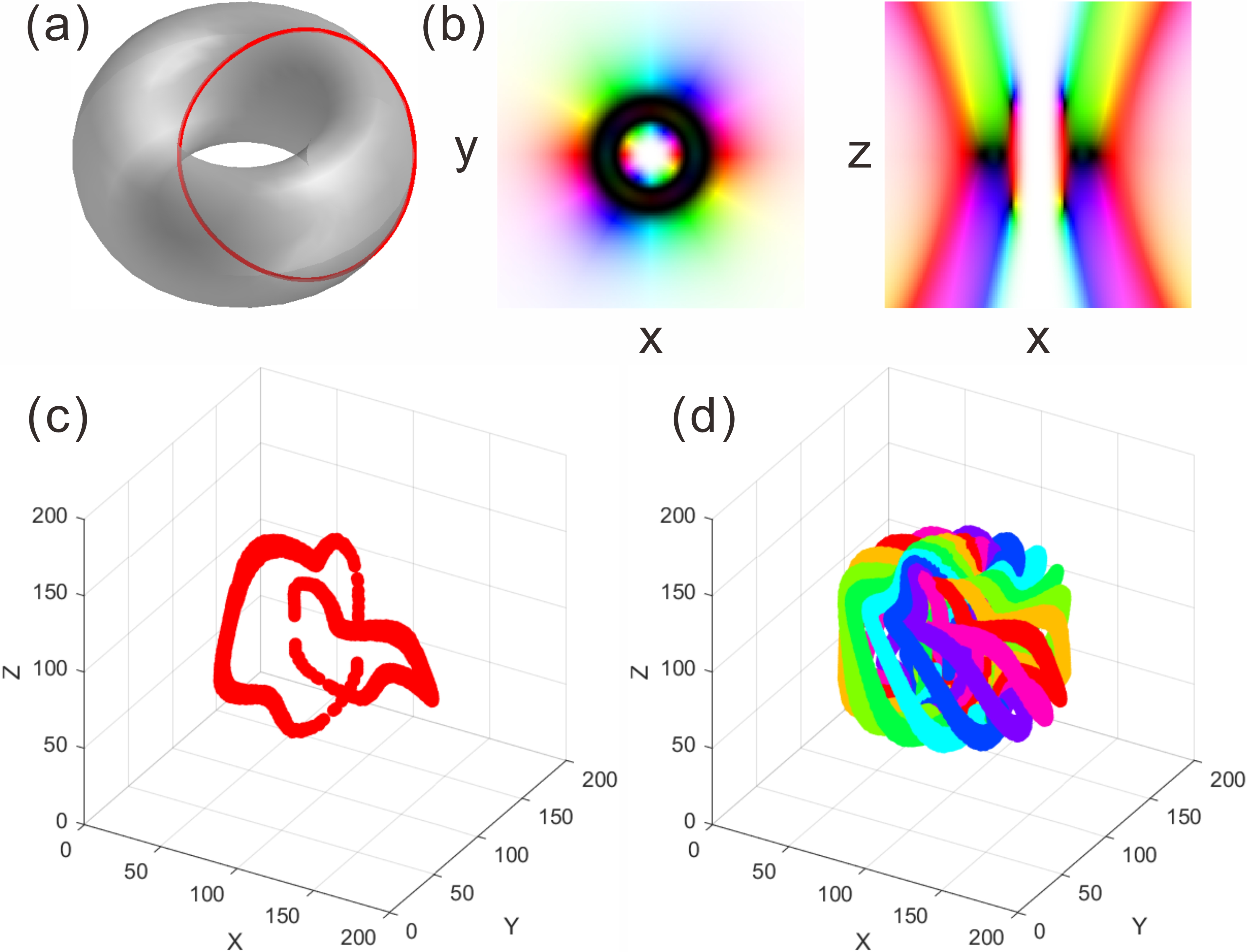}
        \\[0mm] 
        \caption{Schematic of a hopfion with winding numbers $(p, q) = (2, 2)$.  (a), a $(2, 2)$ torus knot obtained from the parametric equations. (b), Polarization distributions in the xy and xz planes.  
(c), isopolar filaments corresponding to the Stokes vector $\mathbf{S} = (1, 0, 0)$.  
(d), a ring-like structure formed by several filaments satisfying $S_z=0$.
} 
        \label{figs3}
    \end{figure*}

As stated in the main text, a single, connected, closed curve of uniform polarization in 3D space exists only when the two winding numbers are coprime.  When this condition is violated, the torus knot factorizes into several disjoint closed curves.  Fig.S\ref{figs3} (a) illustrates this for $(p,q)=(2,2)$: the parametric plot yields a curve indistinguishable from the $(1,1)$ torus knot, precisely because the integers are not coprime.  The $(2,2)$ configuration therefore corresponds to two coincident $(1,1)$ knots; only one of them is visible in the parametric plot, which lies outside the formal range of validity for non-coprime indices.

The LCP and RCP components of a hopfion carrying such winding numbers can be expressed as the coherent superposition of LG modes:
\begin{equation}
    \begin{split}
        \mathbf{E_L}&=\alpha \mathrm{LG}_2^0\mathbf{e_L},\\
        \mathbf{E_R}&=[(1-\beta)(\mathrm{LG}_0^2-\mathrm{LG}_0^1)+\beta \mathrm{LG}_0^0]\mathbf{e_R},
    \end{split}
\end{equation}
where $\alpha=0.5$, $\beta=0.3$ here. Fig.S\ref{figs3} (b) displays the polarization distributions in the $xy$ and $xz$ planes obtained from this expression.  Because $p = 2$, the $xy$ plane hosts a second-order skyrmion, while $q = 2$ gives rise to a pair of first-order skyrmions on one side of the $xz$ plane. By tracing the 3D trajectory of a fixed polarization state—e.g. $\mathbf{S}= (1, 0, 0)$—we obtain two split, mutually linked torus knots, as shown in Fig.S\ref{figs3} (c). Although no single closed curve is formed, each of the two $(1, 1)$ torus knots winds once around the $xy$ origin and pierces the $xz$ mid-plane once, so that their combined topology realizes the total winding numbers $(p, q) = (2, 2)$.  Plotting additional filaments that satisfy $S_z = 0$ yields Fig.S\ref{figs3} (d); even though the winding numbers are not coprime, all polarization states with a fixed $S_z$ still assemble into a toroidal configuration that obeys the Hopf map.  Consequently, our results remain valid, in the general sense, for any non-zero winding numbers.

%